\documentclass[11pt]{article}

\usepackage{amsmath,amssymb}
\usepackage{graphicx}
\usepackage{psfrag}

\newcommand{\R}{\mathbb{R}}
\newcommand{\C}{\mathbb{C}}

\newcommand{\g}{\gamma}
\newcommand{\A}{{\cal A}}

\newcommand{\Abar}{\overline{\cal A}}

\newcommand{\h}{{\cal H}}
\newcommand{\scal}[2]{\langle #1| #2\rangle}

\newcommand{\cyl}{{\rm Cyl}}
\newcommand{\Cyl}{{\rm Cyl}}

\newcommand{\scripta}{\mathfrak{A}}
\newcommand{\ctheta}{\check{\theta}}
\newcommand{\ttheta}{\tilde{\theta}}
\newcommand{\pr}{{\rm pr}}

\usepackage{color}

\newcounter{mnotecount}[section]

\newtheorem{lm}{Lemma}
\newtheorem{df}{Definition}
\newtheorem{thr}{Theorem}

\newtheorem{chr}{Characterization}

\newtheorem{eg}{Example}[section]

\def\be#1\ee{\begin{equation}#1\end{equation}}

\numberwithin{equation}{section}
\numberwithin{lm}{section}
\numberwithin{thr}{section}
\numberwithin{chr}{section}
\numberwithin{df}{section}

\begin{document}

\title{Automorphism covariant representations of the holonomy-flux
$*$-algebra}
\author{ Andrzej Oko{\l}\'ow${}^{1*}$
 and Jerzy\ Lewandowski${}^{1,2\dagger}$}
\date{June 14, 2003}

\maketitle
\begin{center}
{\it 1. Instytut  Fizyki Teoretycznej, Uniwersytet
Warszawski, ul. Ho\.{z}a 69, 00-681 Warsaw, Poland\\
2. Center for Gravitational Physics and Geometry, Physics
Department, 104 Davey, Penn State, University
Park, PA 16802, USA\\
${}^*$ oko@fuw.edu.pl\\
${}^\dagger$ lewand@fuw.edu.pl}
\end{center}
\medskip

\begin{abstract}
We continue the analysis of representations of cylindrical
functions and fluxes which are commonly used as elementary
variables of Loop Quantum Gravity. We consider an arbitrary
principal bundle of a compact connected structure group and,
following Sahlmann's ideas \cite{sahl-1}, 
 define a holonomy-flux $*$-algebra whose elements correspond
 to the elementary variables. There exists a natural action of
 automorphisms of the bundle on the algebra; this
 action generalizes the action of analytic diffeomorphisms and
 gauge transformations on the algebra considered in earlier
 works. We define the automorphism covariance of a
 $*$-representation of the algebra on a Hilbert space and prove
 that the only Hilbert space admitting such a representation is
 a direct sum of the spaces $L^2$, given by a unique
 measure on the space of generalized connections. 
This result is a generalization of our previous work \cite{ol} 
 where we assumed that the principal bundle is trivial and its base manifold is $\R^d$.


\end{abstract}


\section{Introduction}

Loop Quantum Gravity\footnote{For a comprehensive introduction and
references to the subject see \cite{rev}.} (LQG) is an attempt to
obtain a consistent theory of quantum gravity quantizing the
classical theory of general relativity (GR). The departure point
of LQG is a Hamiltonian formulation of GR whose configuration
space is the space of connections on an $SU(2)$ principal fibre
bundle over three-dimensional 'spatial' manifold. The passage from
the classical theory to the quantum theory is done by  canonical
quantization which is required to preserve symmetries of the
Hamiltonian formulation of GR. This means, in particular, that
the quantization  should not make use of any geometrical
background structure on the manifold, e.g. a fixed metric.

A crucial step of the canonical quantization procedure is the
choice of  elementary variables which are (complex) functions on
the phase space of the theory (see e.g. \cite{lqg}).
Once such a set is chosen, the quantization procedure consists
in assigning to any elementary variable an operator on a
Hilbert space in such a way that $(i)$ $(i\hbar)^{-1}$ times the commutator the two operators assigned to a pair of variables corresponds to the Poisson bracket of these variables  and $(ii)$ the conjugate of the operator assigned to a variable corresponds to the complex conjugate of this variable --- if $\hat{f}, \hat{g}$ are the operators assigned respectively to variables $f,g$  then
\[
(i) \ \ (i\hbar)^{-1}[\hat{f},\hat{g}]=\widehat{\{f,g\}}; \ \ \ \ \ \ (ii) \ \ \hat{f}^*=\hat{\bar{f}}.
\]
The assignment $f\mapsto\hat{f}$ is called a representation of elementary variables.

The present paper concerns the theory of representations of the
commonly used elementary variables of LQG which are cylindrical
functions and fluxes. Each (classical) cylindrical function
depends on the holonomies of the $SU(2)$-connections along a
finite number of paths and each (classical) flux is given by
an integral of the momentum variable canonically conjugate to
the connection  over a surface embedded in the 'spatial'
manifold (the momentum can be naturally viewed as a
differential two-form) \cite{ai,area} (for a comprehensive description of the algebra of elementary variables see \cite{acz}). Because of the gauge
and diffeomorphism invariance of LQG any admissible
representation of the variables has to be {\em covariant}
(in an appropriate sense) with respect to the gauge
transformations and the action of diffeomorphisms. By now one
knows only one non-trivial admissible representation of the
variables, hence there arises a question  about the
{\em uniqueness} of the representation.

The question was addressed  for the first time by Sahlmann
\cite{sahl-1, sahl-2} who also proposed a new, more algebraic point
of view on the issue. In Sahlmann's approach one defines a $*$ -algebra
$\scripta$ called a holonomy-flux  $*$-algebra whose elements
correspond to cylindrical functions and fluxes, and then studies
$*$-representations of the algebra on a Hilbert space.  The
space of cylindrical functions used to construct $\scripta$ can be
completed to a $C^*$-algebra $\cyl$  whose Gelfand spectrum is
naturally viewed as the space $\Abar$  of so called generalized
connections \cite{ai,al-difgeom}. The first result by Sahlmann
is that each representation of the holonomy-flux $*$-algebra on a
Hilbert space induces a family of measures on $\Abar$
\cite{sahl-1}. In \cite{sahl-2} Sahlmann studied the
diffeomorphism covariance of $*$-representations of the algebra
$\scripta$  built over $U(1)$-connections. He proved that {\em
every measure given by such a representation coincides with the 
measure introduced  on $\Abar$ by Ashtekar and Lewandowski}
 \cite{al-meas}. This
result was generalized in \cite{sahlthiem} and \cite{ol} to the
case of a holonomy-flux $*$-algebra built over connections with
any  compact connected structure group.

In our earlier work \cite{ol}, we assumed that the holonomy-flux $*$-algebra is built over a {\em trivial} principal bundle over $\R^d$ $(d\geq2)$ as a base ('spatial') manifold. The goal of the present paper is to generalize the results of \cite{ol} to the case of a not necessarily trivializable bundle over an arbitrary base manifold. Thus in the sequel we will consider a principal bundle $P(\Sigma,G)$ such that $(i)$ the base ('spatial') manifold $\Sigma$ is real-analytic\footnote{So far we are not able to construct a holonomy-flux $*$-algebra in the case of any smooth base manifold $\Sigma$. The reason is that the operators generating the algebra are associated with some curves in $\Sigma$ and with some $(d-1)$-dimensional submanifolds of $\Sigma$ $(d=\dim\Sigma)$ and the composition of the operators, i.e. the algebra multiplication, depends on the way in which a given curve intersects a given submanifold. Smooth curves and smooth submanifolds (unlike analytic ones) can intersect each other in a very complicated way which is an obstacle for defining the algebra multiplication in the case of any smooth manifold.} and its dimension $d$ is greater than $1$ and $(ii)$ the structure Lie group $G$ is compact and connected. We will define a holonomy-flux $*$-algebra $\scripta$ associated with such a bundle and will call it the Sahlmann algebra. Next, we will have to define an action of diffeomorphisms of $\Sigma$ and gauge transformations on the algebra. However, in general there is no canonical action of the diffeomorphisms on the space of connections and, consequently, on the algebra $\scripta$. Thus instead of diffeomorphisms and gauge transformations treated separately we will use  {\em automorphisms} of the bundle which unify the notion of diffeomorphisms and gauge transformations. Consequently, we will consider a natural action of the automorphisms on $\scripta$ and define the notion of the {\em automorphism covariance} of a $*$-representation of $\scripta$ on a Hilbert space (precisely speaking, we will consider only those smooth automorphisms of $P(\Sigma,G)$ which define {\em analytic} diffeomorphisms on the base manifold $\Sigma$). As a main result of the paper, we will show that the only measure on $\Abar$ which admits an automorphism covariant $*$-representation of the Sahlmann algebra is the Ashtekar-Lewandowski measure.

In this paper we are able to prove merely the uniqueness of the measure on $\Abar$ which admits an automorphism covariant representation of the Sahlmann algebra, while the question of the uniqueness of the representation was left open. However, one can go a step further and prove the uniqueness of an automorphism covariant representation of the algebra which will be shown in \cite{lost}.

Let us finally emphasize that in the present paper we will skip
the 'derivation' of the Sahl\-mann algebra from the elementary
variables, i.e. we will just give the definition of the algebra.
 The reader can find such a derivation, well suited for the
 considerations presented in the sequel, in \cite{ol}.

\section{Preliminaries}

Let $\Sigma$ be a real-analytic $d$-dimensional manifold ($d\geq 2$) and let  $P(\Sigma,G)$ be a principal bundle over $\Sigma$, whose structure Lie group $G$ is compact and connected. The right action of the group on $P$ will be denoted by $R_g$:
\[
P\times G \ni(p,g)\mapsto R_g(p)\in P,
\]
and $\rm pr$ will denote the projection from $P$ onto $\Sigma$ defined by $R_g$.

Let $M$ be any (possibly 0-dimensional) submanifold in $\Sigma$ and let $P_M$ denote the restriction of the bundle $P$ to $M$. We say that a map $\theta:P_M\rightarrow P$ is a smooth morphism on $P_M$ if $(i)$ $\theta$  is an invertible smooth map and $\theta^{-1}$  is smooth and $(ii)$ $\theta$ commutes with the right action $R_g$ for every $g\in G$. The morphism $\theta$ is an automorphism of a bundle $P_M$ if $\theta(P_M)=P_M$. Denote by $\Theta$ the set of all smooth automorphisms of $P$.

Every automorphism $\theta\in \Theta$ defines a smooth diffeomorphism $\ctheta$ on $\Sigma$ such that:
\[
\ctheta\circ\pr=\pr\circ\theta.
\]
The automorphism $\theta$ is {\em vertical} if and only if $\ctheta={\rm id}$. Clearly, the set  of all vertical automorphisms is a subgroup of $\Theta$. Another subgroup of $\Theta$ is the one containing all automorphisms such that they define {\em analytic} diffeomorphisms on $\Sigma$:
\[
\Theta^\omega:=\{\theta \ | \ \theta\in\Theta \ \text{and $\check{\theta}$ is analytic}\}.
\]

Given an automorphism $\theta\in\Theta$ and a morphism $\theta'$ on $P_M$ we define a morphism ${\rm ad}_\theta\theta'$ on $P_{\check{\theta}(M)}$ in the following way:
\begin{equation}
P_{\check{\theta}(M)}\ni p\mapsto {\rm ad}_\theta(\theta'(p)):=\theta(\theta'(\theta^{-1}(p)))\ \in P.
\label{ad}
\end{equation}

Let $\A$ be a set of all smooth connections on $P$. We will describe every element of $\A$ by means of a connection one-form $A$ on $P$ valued in the Lie algebra $G'$ of the group $G$. Usually, one describes a connection by a (family of) one-form(s) on the base manifold $\Sigma$, which is the pull-back of $A$ with respect to a (family of local) section(s) of $P$. However, an arbitrary bundle $P$ does not possess any distinguished section---therefore we will avoid the latter way of describing connections.

\section{Sahlmann algebra}

The Sahlmann algebra is an algebra of operators acting on a linear space of some complex functions defined on $\A$. Operators constituting the algebra are associated with some submanifolds embedded in $\Sigma$. We impose some regularity conditions on the submanifolds underlying the algebra and call them  regular analytic submanifolds \cite{ol}:

\begin{df}
$(i)$  $0$-dimensional regular analytic submanifold of $\Sigma$ is a one-element subset of $\Sigma$; $(ii)$ an $m$-dimensional $(0<m<\dim\Sigma)$ regular analytic submanifold $M$ is an analytic orientable $m$-dimensional submanifold such that its closure $\overline{M}$ is a compact subset of an analytic $m$-dimensional submanifold of $\Sigma$ and $\overline{M}\setminus M$ is a finite union of regular analytic submanifolds of lower dimensions.
\end{df}
The following subsets of $\Sigma$ will serve to define the space of functions, on which Sahlmann algebra acts.
\begin{df}
$(i)$ An analytic oriented edge in $\Sigma$ is the closure of any connected $1$-dimensional regular analytic submanifold with fixed orientation.
$(ii)$ An analytically embedded graph $\gamma$ in $\Sigma$ is a finite set of
analytic oriented edges in $\Sigma$ such that two
distinct edges can intersect each other only at their endpoints. The
edges  will be called the edges of the graph $\gamma$, and their endpoints --- the vertices of $\gamma$.
\end{df}

\subsection{Cylindrical functions \label{cyl-funs}}

In this section we will introduce the notion of cylindrical function following \cite{B1}.

Let $e$ be an oriented edge. The orientation of $e$  allows us to call one of its endpoints the source and the other one---the target of $e$. We will denote them by $(e)_s$ and $(e)_t$ respectively.  Each connection $A$ defines a parallel transport $A(e)$ along $e$ being an isomorphism from $P_{(e)_s}$ onto $P_{(e)_t}$, which commutes with the right action $R_g$ for every $g\in G$; this means  that $A(e)$ is a morphism on $P_{(e)_s}$ of image $P_{(e)_t}$. We also have:
\begin{equation}
A(e_1\circ e_2)=A(e_1)\circ A(e_2),\;\;\;\;A(e^{-1})=A^{-1}(e),
\label{properties-A}
\end{equation}
where $e_1\circ e_2$ denotes the composition of two edges such that $(e_1)_s=(e_2)_t$ and $e^{-1}$ (the inverse of $e$) is an edge obtained from $e$ by the change of the orientation.

Consider now a graph $\gamma$ constituted by edges $\{e_1,\ldots,e_N\}$ and define an equivalence relation on $\A$ in the following way. Given $A,A'\in\A$
\begin{equation}
A\sim_\g A'\ \ \text{if and only if}\ \ A(e_I)=A'(e_I)
\label{rel}
\end{equation}
for every $I\in\{1,\ldots,N\}$. Denote by $A_\g$ the equivalence class of a connection $A$ and by $\A_\gamma$---the set of all equivalence classes defined by the relation. The relation $\sim_\g$ defines the following projection:
\[
\A \ni A\mapsto p_\g(A):=A_\g \in\A_\g.
\]

An important fact is that graphs in $\Sigma$ form a directed set with relation $\geq$ defined as follows: $\gamma'\geq \gamma$ if and only if $(i)$ every edge of $\gamma$ can be expressed as a composition of edges of $\gamma'$ and their inverses and $(ii)$ each vertex of $\gamma$ is a vertex of $\gamma'$ \cite{al-difgeom}. It follows from (\ref{properties-A}) that if $\gamma'\geq\gamma$ then there exists a surjective projection $p_{\gamma\gamma'}:\A_{\gamma'}\rightarrow\A_\gamma$ such
that:
\begin{equation}
p_\gamma= p_{\gamma\gamma'}\circ p_{\gamma'}.
\label{proj-map}
\end{equation}

Let $P_{V(\g)}$ be the bundle obtained by the restriction of the bundle $P$ to the set $V(\g)$ of the vertices of $\g$. Then every $A_\g$ can be thought as an $N$-tuplet of morphisms 
\[
(A(e_1),\ldots, A(e_N))=(\,A(e_I)\,)
\]
acting on appropriate fibres of $P_{V(\g)}$. The bundle  $P_{V(\g)}$ is trivializable; every trivialization of $P_{V(\g)}$ defines a bijection from $\A_\g$ onto $G^N$ \cite{al-meas} and hence a structure of a (real-analytic) manifold on $\A_\g$. It turns out that this structure does not depend on the choice of the trivialization. The projection $p_{\gamma\gamma'}$ mentioned above is an analytic map from $\A_{\gamma'}$ onto $\A_\gamma$.

\begin{df}
We say that $\Psi:\A\rightarrow\C$ is a cylindrical function compatible with a graph $\gamma$, iff:
\be
\Psi\ =\ p_\g^*\psi.
\label{cyl}
\ee
where $\psi$ is a complex function on $\A_\g$.
\end{df}

Each cylindrical function is compatible with many graphs: e.g. if $\Psi$  is compatible with $\g$ and if $\g'\geq \g$, then $\Psi$ is compatible with $\g'$ as well, which follows from (\ref{proj-map}).

If the function $\psi:\A_\g\rightarrow\C$ in (\ref{cyl}) is smooth we say that the corresponding cylindrical function $\Psi:\A\rightarrow\C$ is smooth as well. Let
\[
\cyl^\infty:={\rm span}\ \{\ \Psi \ | \ \text{$\Psi$ is a smooth cylindrical function on $\A$}\}.
\]
In fact, $\cyl^\infty$ is not only a linear space, but an algebra also. To see this, note first that every element $\Psi'$ of the space is a finite linear combination of smooth cylindrical functions $\{\Psi_i\}$ $(i=1,\ldots,k)$ compatible, respectively, with  graphs  $\{\g_i\}$. There exists a graph $\g$ such that $\g\geq\g_i$ for every $i$ under consideration which means that every function $\Psi_i$ is compatible with the graph. Hence $\Psi'$ is a smooth cylindrical function compatible with $\g$ also. Now, given elements $\Psi$ and $\Psi'$ of $\cyl^\infty$ we can find a graph $\g'$ such that the two functions are compatible with it. Then $\Psi\Psi'$ is a smooth cylindrical function compatible with $\g'$, thus an element of $\cyl^\infty$.  

\subsection{Flux operators\label{flux-sec}}

Let us now define some differential operators on $\cyl^\infty$. Let $S$ be a $(d-1)$-dimensional regular analytic manifold with fixed (external) orientation and $P_S$ be the restriction of the bundle $P$ to $S$. Denote by $\lambda$ a smooth map $\R\times P_S\rightarrow P_S$ such that for each arbitrary but fixed $\tau$ the map $\lambda_\tau:=\lambda(\tau,\cdot)$ is a vertical automorphism of $P_S$ such that:
\[
\lambda_{\tau+\tau'}=\lambda_\tau\circ\lambda_{\tau'}\ \text{and}\ \lambda_0={\rm id},
\]
i.e. $\lambda_\tau$ is a one-parameter group of vertical automorphisms of $P_S$.

To define the differential operators on $\cyl^\infty$  we distinguish graphs {\em a\-dap\-ted} to $S$---we say that a graph $\gamma$ is adapted to $S$ iff every edge of $\gamma$ either: $(i)$ is contained  in $S$ (modulo its endpoints), or $(ii)$ does not intersect $S$ at all, or $(iii)$ intersects $S$ at exactly one  endpoint. Next, among edges belonging to the class $(iii)$ we distinguish ones placed `up' (`down') the {\em oriented} submanifold $S$. Note that if a graph $\gamma'$ is not adapted to $S$, then, thanks to the analyticity of $S$ and the edges of $\g'$, one can  adapt the graph to the submanifold by the appropriate subdividing of its edges i.e. for every $\g'$ there exists $\g\geq\g'$ such that $\g$ is adapted to $S$ \cite{area}. Thus every cylindrical function is compatible with a graph adapted to a given $S$.

Suppose now that a graph $\gamma$ is adapted to $S$.  
By means of a given family $\lambda_\tau$ of vertical automorphisms of $P_S$ we define a family of smooth maps\footnote{The maps $\boldsymbol{\lambda}_\tau$ ($\tau\in\R)$ define smooth curves on $\A_\g$, which cannot be lifted to curves on $\A$. But as it was noted by Professor S.L. Woronowicz the maps do define curves on the space $\Abar$ of generalized connections (for the definition of $\Abar$ see Footnote \ref{gen-con}).} $\boldsymbol{\lambda}_\tau: \A_\g\rightarrow\A_\g$ by the formula
\begin{equation}
\boldsymbol{\lambda}_{\tau}(A_\g):=(\lambda^{I_t}_{\tau}A(e_I)\lambda^{I_s}_{\tau}),
\label{curve}
\end{equation}
where $\lambda^{I_s}_\tau$ and $\lambda^{I_t}_\tau$ are automorphisms of the fibres $P_{(e_I)_s}$ and $P_{(e_I)_t}$ respectively defined as follows:
\begin{enumerate}
\item if the edge $e_I$ belongs to the class $(i)$ or $(ii)$ described above, then $\lambda^{I_t}_{\tau}={\rm id}$ and $\lambda^{I_s}_{\tau}={\rm id}$;
\item if the edge $e_I$ belongs to the class $(iii)$ and
\begin{enumerate}
\item $S\cap e_I=(e_I)_s$, i.e. $e_I$ is 'outgoing' from $S$ then  $\lambda^{I_s}_{\tau}$ acts on the fibre $P_{(e_I)_s}$ as
\begin{enumerate}
\item $\lambda_\tau$ if $e_I$ is placed 'up' the submanifold $S$,
\item $\lambda_{-\tau}$ if $e_I$ is placed 'down' the submanifold $S$,
\end{enumerate}
and $\lambda^{I_t}_{\tau}={\rm id}$.
\item if $S\cap e_I=(e_I)_t$ i.e. $e_I$ is 'ingoing' to  $S$ then $\lambda^{I_t}_{\tau}$ acts on the fibre $P_{(e_I)_t}$ as
\begin{enumerate}
\item $\lambda_{-\tau}$ if $e_I$ is placed 'up' the submanifold $S$,
\item $\lambda_{\tau}$ if $e_I$ is placed 'down' the submanifold $S$,
\end{enumerate}
and $\lambda^{I_s}_{\tau}={\rm id}$.
\end{enumerate}
\end{enumerate}

Let $\Psi=p_\g^*\psi$ be a smooth cylindrical function compatible with the graph $\g$ adapted to $S$.

\begin{df}
Flux operator $\hat{X}_{S,\lambda}$ associated with the submanifold $S$ and the family $\lambda_\tau$ of vertical automorphisms acts on $\Psi$ in the following way:
\[
\hat{X}_{S,\lambda}\Psi:=-\frac{i}{2}\frac{d}{d\tau}\Big|_{\tau=0}(\ p^*_\g\ \boldsymbol{\lambda}^*_\tau\ \psi\ ).
\]
\label{flux-df}
\end{df}
Although the function $\Psi$ is compatible with many distinct graphs adapted to $S$ the r.h.s. of the above equation does not depend on the choice of an adapted graph. Clearly, $\hat{X}_{S,\lambda}$ is a linear operator defined on the whole $\cyl^\infty$ and preserves the space. It is easy to see that in the case of a trivial bundle Definition \eqref{flux-df} is equivalent to the standard definition of flux operators introduced in \cite{area}.

For $S,S_1,S_2$ such that $S=S_1\cup S_2$ and $S_1\cap S_2=\varnothing$ and for any $\lambda_\tau$ defined on $P_S$  we have:
\begin{equation}
\hat{X}_{S,\lambda}=\hat{X}_{S_1,\lambda_1}+\hat{X}_{S_2,\lambda_2},
\label{sum-X}
\end{equation}
where $(\lambda_i)_\tau:=\lambda_\tau|_{P_{S_i}}$ ($i=1,2$).

\subsection{Definition of the Sahlmann algebra}

Let $\scripta$ be a complex algebra generated by all operators on $\cyl^\infty$ of the following form:
\begin{equation}
\begin{gathered}
\Psi\mapsto \hat{\Phi} \Psi:=\Phi\Psi,\\
\Psi\mapsto \hat{X}_{S,\lambda} \Psi,
\end{gathered}
\label{oper}
\end{equation}
where $\Psi,\Phi\in\cyl^\infty$. Define $*$ operation on $\scripta$:
\[
\hat{\Phi}^*:=\hat{\overline{\Phi}},\;\;\;\hat{X}_{S,\lambda}^*:=\hat{X}_{S,\lambda}.
\]
To show that the $*$ operation is well defined on $\scripta$ note first that the above formulas define a $*$ operation on the free algebra $\scripta_F$ generated by all the smooth cylindrical functions and the flux operators. Obviously, every element $a$ of $\scripta_F$ defines a linear operator on $\cyl^\infty$ via \eqref{oper}, which will be denoted by $a$ also. Then
\[
\scripta=\scripta_F/\scripta_0,
\]    
where $\scripta_0$ is a (left and right) ideal in $\scripta_F$ defined as follows:
\[
\scripta_0:=\{a\in\scripta_F \ | \ a\Psi=0 \ \text{for all} \ \Psi\in\cyl^\infty\}.
\]  
Now, the $*$ operation is well defined on $\scripta$ if 
\begin{equation}
a\in\scripta_0 \Longrightarrow a^*\in\scripta_0. 
\label{impl}
\end{equation}

To prove the implication fix an element $a$ of $\scripta_F$ and a smooth cylindrical function $\Psi$. The element $a$ is generated by a finite number of smooth cylindrical functions and flux operators. It is possible to find a graph $\g$ such that $(i)$ $\g$ is adapted to all the surfaces corresponding to the flux operators, $(ii)$ all the cylindrical functions are compatible with $\g$ and $(iii)$ the function $\Psi$ is compatible with $\g$. Now it is easy to see that $a\Psi$ is also compatible with $\g$.  

Consider now the space $\A_\g$. As it was mentioned earlier, any trivialization of the bundle $P_{V(\g)}$ defines a bijection between $\A_\g$ and $G^N$, where $N$ is the number of the edges of $\g$. This bijection can be used to push-forward the Haar measure on $G^N$ onto $\A_\g$ \cite{al-meas}. The resulting measure $d\mu_\g$ is independent of the choice of the trivialization of $P_{V(\g)}$ and can be used to define a scalar product $\scal{\cdot}{\cdot}_\g$ on the linear space of smooth cylindrical functions compatible with the graph $\g$. Then we have \cite{area}
\[
\scal{\Psi}{a\Psi'}_\g=\scal{a^*\Psi}{\Psi'}_\g
\]                         
for every pair $\Psi,\Psi'$ of smooth cylindrical functions compatible with $\g$. Thus if $a\in\scripta_0$ then 
\[
\scal{a^*\Psi}{\Psi'}_\g=0.
\]
Taking into account smoothness of $a^*\Psi$ we conclude that $a^*\Psi=0$, which proves the implication \eqref{impl}.  

Now we are able to give a definition of the Sahlman algebra:

\begin{df}
Sahlmann holonomy-flux $*$-algebra (Sahlmann algebra for short) is the $*$-algebra $(\scripta,*)$ \cite{sahl-1,ol}.
\end{df}

We emphasize that the Sahlmann algebra is a unital one with a unit given by a constant cylindrical function of the value equal to $1$.

\section{Induced action of automorphisms on $\scripta$}

There is a natural (right) action of the group of automorphisms of $P$ on the space $\A$:
\begin{equation}
\A\ni A\mapsto \theta^*A\in\A,
\label{pull-back}
\end{equation}
where $\theta\in\Theta$. Given $\theta\in\Theta^\omega$ and $A\in\A$ the parallel transport $(\theta^*A)(e):P_{e_s}\rightarrow P_{e_t}$ can be expressed as
\begin{equation}
(\theta^*A)(e)={\rm ad}_{\theta^{-1}} (A(\ctheta(e))).
\label{theta-A}
\end{equation}
This suggests to define a map\footnote{The map ${\rm Ad}_{\theta}$ is an analytic diffeomorphism, what can be easily seen after expressing the map in any trivializations of bundles $P_{V(\g)}$ and $P_{V(\ctheta(\g))}$.}:
\[
\A_\g\ni A_\g\mapsto {\rm Ad}_\theta A_\g:=(\ {\rm ad}_\theta \, (A(e_I))\ )\in \A_{\ctheta(\g)}.
\]
where edges $\{e_I\}$ constitute the graph $\g$. Applying (\ref{theta-A}) to the edges we obtain
\[
p_\g \circ \theta^* = {\rm Ad}_{\theta^{-1}}\circ p_{\ctheta(\g)},
\]
where the maps on both sides of the latter equation are from $\A$ onto $\A_\g$.

The right action (\ref{pull-back}) induces a linear left action of $\theta$ (that is, a representation of the group $\Theta$)  on the space of functions on $\A$:
\[
(\ttheta\Psi)(A):=\Psi(\theta^*A),
\]
where $\Psi$ is a function on $\A$. Let $\Psi_n$ be a sequence of functions on $\A$ converging pointwisely to a function $\Psi$. Then
\begin{equation}
\lim_{n\rightarrow\infty}\ttheta\Psi_n=\ttheta(\lim_{n\rightarrow\infty}\Psi_n)=\ttheta\Psi,
\label{theta-cont}
\end{equation}
which means that $\ttheta$ is continuous with respect to the topology of pointwise convergence.

We have the following lemma:

\begin{lm}
If $\theta\in\Theta^\omega$  and $\Psi\in\cyl^\infty$ then $\ttheta\Psi\in\cyl^\infty$.
\end{lm}

\noindent {\bf Proof.} Let $\Psi\in\Cyl^\infty$ be compatible with the graph $\gamma$. Then
\begin{multline*}
(\ttheta\Psi)(A)=\Psi(\theta^*A)=\psi(p_\g(\theta^*A))=\psi({\rm Ad}_{\theta^{-1}}p_{\ctheta(\g)}(A))=\\=[p_{\ctheta(\g)}^*({\rm Ad}^*_{\theta^{-1}}\psi)](A),
\end{multline*}
that is:
\begin{equation}
\ttheta\Psi=p_{\ctheta(\g)}^*({\rm Ad}^*_{\theta^{-1}}\psi),
\label{theta-action}
\end{equation}
where ${\rm Ad}^*_{\theta^{-1}}\psi$ is a smooth complex function on $\A_{\ctheta(\gamma)}$. Thus $\ttheta\Psi$ is a smooth cylindrical function compatible with the (analytic) graph $\ctheta(\g)$. $\blacksquare$

Now we can define the action of the automorphism $\theta$ on the Sahlmann algebra:
\begin{equation}
\scripta\ni \hat{a}\mapsto \ttheta \hat{a}\ttheta^{-1}.
\label{th-act-Sh}
\end{equation}
\begin{lm}
The map (\ref{th-act-Sh}) is  a $*$-preserving isomorphism of the algebra $\scripta$ onto itself.
\end{lm}
The above lemma is an implication of the following one:
\begin{lm}
(i) If $\Phi\in\cyl^\infty$, then:
\[
\ttheta \hat{\Phi}\ttheta^{-1}=\widehat{\ttheta\Phi}.
\]
(ii) For flux operators the following formula is true:
\[
\ttheta \hat{X}_{S,\lambda}\ttheta^{-1}=\hat{X}_{\check{\theta}(S),{\rm ad}_\theta\lambda},
\]
where $({\rm ad}_\theta\lambda)_\tau:={\rm ad}_\theta\lambda_\tau$.
\label{lm-transf}
\end{lm}

\noindent {\bf Proof} of the statement $(ii)$ (the proof of $(i)$ is trivial). We have:
\begin{multline}
\frac{2}{i}\ \ttheta\ (\hat{X}_{S,\lambda}\Psi)\ =\ \ttheta\ [\ \frac{d}{d\tau}\Big|_{\tau=0}\ (\ p^*_\g\ \boldsymbol{\lambda}^*_\tau\ \psi\ )\ ]\ =\\=\ \frac{d}{d\tau}\Big|_{\tau=0}\ [\ \ttheta \ (p^*_\g\,\boldsymbol{\lambda}^*_\tau\,\psi)\ ]\ =\ \frac{d}{d\tau}\Big|_{\tau=0}\ [\ p^*_{\ctheta(\g)}\ ({\rm Ad}^*_{\theta^{-1}} \boldsymbol{\lambda}^*_\tau \psi)\ ]\ =\\=
\frac{d}{d\tau}\Big|_{\tau=0}\ [\ p^*_{\ctheta(\g)}\ ({\rm Ad}^*_{\theta^{-1}}\, \boldsymbol{\lambda}^*_\tau \,{\rm Ad}^*_{\theta})\ ({\rm Ad}^*_{\theta^{-1}}\psi)\ ].
\label{theta-X}
\end{multline}
In the second step of above calculation we used the linearity of $\ttheta$ and (\ref{theta-cont}) to change the order of the action of $\ttheta$ and the differentiation with respect to $\tau$, in the third one we applied (\ref{theta-action}) to the function $p^*_\g(\boldsymbol{\lambda}^*_\tau\psi)$ as cylindrical one compatible with the graph $\g$.

Definition (\ref{curve}) and Equation (\ref{ad}) imply that the family of maps ${\rm Ad}_{\theta}\circ \boldsymbol{\lambda}_\tau \circ {\rm Ad}_{\theta^{-1}}$ acting on (and preserving) the space $\A_{\check{\theta}(\g)}$ is defined by the family $({\rm ad}_\theta\lambda)_\tau$ of vertical automorphisms of $P_{\ctheta(S)}$, i.e.
\[
{\rm Ad}_{\theta}\circ \boldsymbol{\lambda}_\tau \circ {\rm Ad}_{\theta^{-1}}=({\rm \bf ad}_{\boldsymbol{\theta}}\boldsymbol{\lambda})_\tau.
\]
Setting this to the r.h.s of Equation (\ref{theta-X}) we get:
\[
\ttheta(\hat{X}_{S,\lambda}\Psi)=\hat{X}_{\check{\theta}(S),{\rm ad}_\theta\lambda}(\ttheta\Psi).
\]
$\blacksquare$

\section{Representations of Sahlmann algebra}

The Sahlmann algebra $\scripta$ is not equipped with any norm. This does not allow us to expect that one can represent the algebra by means of {\em bounded} operators on a Hilbert space. The fact leads to the following definition of a $*$-representation of $\scripta$ \cite{ol}:
\begin{df}
Let $L(\h)$ be a space of linear operators on a Hilbert space $\h$. We say that a map $\pi:\scripta\rightarrow L(\h)$ is a $*$-representation of $\scripta$ on the Hilbert space $\h$ if:
\begin{enumerate}
\item there exists a dense subspace ${\cal D}$ of $\h$ such that
\[
{\cal D}\subset\bigcap_{\hat{a}\in\scripta}[\ D(\pi(\hat{a}))\cap D(\pi^*(\hat{a}))\ ],
\]
where $D(\pi(\hat{a}))$ denotes the domain of the operator $\pi(\hat{a})$;

\item for every $\hat{a},\hat{b}\in\scripta$ and $\lambda\in\C$ the following conditions are satisfied on ${\cal D}$:
\begin{alignat*}{3}
\pi(\hat{a}+\hat{b})&=\pi(\hat{a})+\pi(\hat{b}),&\qquad\pi(\lambda\hat{a})&=\lambda\pi(\hat{a}),\\
\pi(\hat{a}\hat{b})&=\pi(\hat{a})\pi(\hat{b}),&\qquad \pi(\hat{a}^*)&=\pi^*(\hat{a}).
\end{alignat*}
\item If, given $\hat{a}\in\scripta$, there exists a subspace $E$ dense in $\h$ such that $E\subset D(\pi(\hat{a}))$ and $\pi(\hat{a})|_E$ is closable, then $\pi(\hat{a})$ is equal to the closure of $\pi(\hat{a})|_E$.
\end{enumerate}
We say that $\pi$ is non-degenerate iff the fact that for every $\hat{a}\in\scripta$ $\pi(\hat{a})v=0$  implies $v=0$.
\label{repr-df}
\end{df}
These conditions mean in particular that $\pi(\hat{a}){\cal D}\subset{\cal D}$ and that every element $\hat{a}=\hat{a}^*$ is represented by the operator $\pi(\hat{a})$ {\em symmetric} on $\cal D$.

In the sequel we will consider only non-degenerate $*$-representations of $\scripta$.

\subsection{Basic facts}

Although Sahlman algebra is not equipped with any norm we can define a norm on its subalgebra $\cyl^\infty$:
\[
\|\hat{\Psi}\|_{\sup}:=\sup_{A\in\A}|\Psi(A)|.
\]
If $\pi$ is a non-degenerate $*$-representation of $\scripta$ on $\h$, then $\pi|_{\cyl^\infty}$ maps elements of $\cyl^\infty$ into {\em bounded} operators on $\h$ and is uniquely extendable to a (non-degenerate) representation of a $C^*$-algebra $\cyl$ defined as the completion of $\cyl^\infty$ in the norm $\|\cdot\|_{\sup}$ \cite{ol}. The connection between the representations $\pi|_{\cyl^\infty}$ and the representations of the $C^*$-algebra $\cyl$ implies the following characterization of $\pi$ and $\h$ \cite{sahl-1}:
\begin{chr}[Sahlmann]\mbox{}\smallskip
\begin{enumerate}
\item The representation $\pi|_{\cyl^\infty}$ is a direct sum of cyclic representations
$\{\pi_\nu\}$:
\be\label{A}
\h=\bigoplus_{\nu\in{\cal N}}\h_\nu,\;\;\;\pi|_{\cyl^\infty}=
\bigoplus_{\nu\in {\cal N}}\pi_\nu,
\ee
where $\{\h_\nu\}$ are carrier spaces of representations $\{\pi_\nu\}$,
 respectively,  $\nu$ ranges some label set ${\cal N}$ and the sum
is orthogonal;
\item For each $\nu$ there exists a Hilbert space isomorphism
\be\label{B}
\varphi_\nu:   L^2(\Abar,\mu_\nu) \  \rightarrow \ \h_\nu,
\ee
where $\Abar$ is a Gel'fand-Neimark spectrum\footnote{There exists the following geometric characterization of the spectrum of $\cyl$. Let $B_{x,y}$ ($x,y\in\Sigma$) be a set of all morphisms from $P_x$ onto $P_y$. The set $B:=\bigcup B_{x,y}$, where $(x,y)$ runs over $\Sigma\times\Sigma$, possesses a natural groupoid structure. A {\em generalized connection} $\bar{A}$ on $P$ is a homomorphism from the groupoid of analytic oriented edges in $\Sigma$ into $B$ such that it maps every oriented edge $e$ to a morphism $\bar{A}(e)\in B_{(e)_s,(e)_t}$. The fact that $\bar{A}$ is a homomorphism means that it satisfies conditions (\ref{properties-A}). The spectrum of $\cyl$  can be identified with the set $\Abar$ of all generalized connections on $P$. Clearly, $\A\subset\Abar$. For more details see \cite{al-difgeom}.\label{gen-con}}
of $\cyl$, and $\mu_\nu$ is a regular, Borel measure on $\Abar$.
\end{enumerate}
\label{dec}
\end{chr}
Every Hilbert space $\h_\nu$ is defined by the choice of a vector $v_\nu\in\h$:
\begin{equation}
\h_\nu:=\overline{\{ \pi(\hat{\Phi})v_\nu \ | \ \Phi\in\cyl^\infty\}},
\label{h-nu}
\end{equation}
the measure $\mu_\nu$ is given by the following formula:
\begin{equation}
\int_{\Abar}\Phi \, d\mu_\nu:=\scal{v_\nu}{\pi(\hat{\Phi})v_\nu},
\label{measure-df}
\end{equation}
and the isomorphism $\varphi_\nu$ is the closure of the map:
\begin{equation}
L^2(\Abar,\mu_\nu)\supset\cyl^\infty\ni\Phi \ \mapsto \  \pi(\hat{\Phi})v_\nu\in\h_\nu
\label{iso}
\end{equation}
(here $\scal{\cdot}{\cdot}$ is the scalar product on $\h$). Clearly, the set of isomorphisms $\{\varphi_\nu\}$ defines an isomorphism $\varphi$, which maps $\bigoplus_{\nu\in{\cal N}}L^2(\Abar,\mu_\nu)$ onto $\h$. In the sequel we will assume that $\varphi$ is fixed and will identify both the Hilbert spaces. Similarly, we will not distinguish between representation $\pi$ on $\h$ and $\varphi^{-1}\pi\varphi$  on  $\bigoplus_{\nu\in{\cal N}}L^2(\Abar,\mu_\nu)$ and both will be denoted by $\pi$.

Given $\Psi\in \h$, we will denote by $\Psi_\nu$ the orthogonal projection
of $\Psi$ onto $\h_\nu=L^2(\Abar,\mu_\nu)$ and write $\Psi=(\Psi_\nu)$. Following \cite{sahl-1} we define
\begin{multline}
{\cal C}^\infty:=\{ (\Psi_\nu)\in \h \ \big|\; \text{$\Psi_\nu=0$ for all but finitely many $\nu$'s}\\ \text{and $\Psi_\nu\in \cyl^\infty$ for every $\nu$}\}.
\label{c-infty}
\end{multline}
Since $\Cyl^\infty$ is dense in $L^2(\Abar,\mu_\nu)$, ${\cal C}^\infty$ is dense in $\h$. It turns out that the assumption ${\cal D}={\cal C}^\infty$, where $\cal D$ is introduced by Definition \ref{repr-df}, gives the following useful characterization \cite{sahl-1}:

\begin{chr}[Sahlmann]
Suppose $\pi$ is a non-degenerate $*$-rep\-re\-sen\-ta\-tion  of $\scripta$ in a Hilbert
space $\h$, and $\pi$ satisfies the assumption ${\cal D}={\cal C}^\infty$.
For every $\hat{X}_{S,\lambda}$,
$\pi$ defines a family of elements of $\h$ labelled by elements of
the set ${\cal N}$ (the same as in Equation (\ref{A}))
\be {\cal
N}\ni\iota\mapsto {F_{S,\lambda}}^\iota\in \h,
\ee
such that the
following conditions are satisfied (given ${F_{S,\lambda}}^\iota$,  we
will subsequently denote by ${{F_{S,\lambda}}^\iota}_\nu$ the $\h_\nu$
component of ${F_{S,\lambda}}^\iota$):

\begin{enumerate}
\item for every $\Psi=(\Psi_\nu)\in{\cal C}^\infty$:
\[
\pi(\hat{X}_{S,\lambda})\Psi=\hat{\mathbf{X}}_{S,\lambda}\Psi+\hat{F}_{S,\lambda}\Psi
\]
where $\hat{\mathbf{X}}_{S,\lambda}\Psi:=(\hat{X}_{S,\lambda}\Psi_\nu)$ and:
\[
\hat{F}_{S,\lambda}\Psi=\hat{F}_{S,\lambda}(\Psi_\nu):=
(\sum_\iota\Psi_{\iota}{F_{S,\lambda}}^{\iota}\!_{\nu}),
\]
where $\sum_\iota\Psi_{\iota}{F_{S,\lambda}}^{\iota}\!_{\nu}$ belongs to $\h_\nu$.  
\item for every $\Phi,\Phi'\in \cyl^\infty\subset \h_\nu$
\begin{equation}
\scal{\hat{X}_{S,\lambda}\Phi}{\Phi'}_\nu-\scal{\Phi}{\hat{X}_{S,\lambda}\Phi'}_\nu=
\scal{\Phi}{({F_{S,\lambda}}^\nu\!_\nu-{\overline{F}_{S,\lambda}}^\nu\!_\nu)\Phi'}_\nu,
\label{div-X}
\end{equation}
where $\scal{\cdot}{\cdot}_\nu$ is the scalar product on $\h_\nu$ 
({\bf \em no} summation with respect to the
index $\nu\in{\cal N}$ in ${F_{S,\lambda}}^\nu\!_\nu$);
\item for every $S=S_1\cup S_2$ such that  $S_i$ ($i=1,2$) are disjoint:
\begin{equation}
{{F_{S,\lambda}}^{\iota}}_\nu={{F_{S_1,\lambda_1}}^{\iota}}_\nu+{{F_{S_2,\lambda_2}}^{\iota}}_\nu,
\label{F-sum}
\end{equation}
where $(\lambda_i)_\tau:=\lambda_\tau|_{P_{S_i}}$.
\end{enumerate}
\label{sahl-th}
\end{chr}

\subsection{Automorphism covariant representations of Sahlmann algebra \label{work}}

Following \cite{ol} we formulate:
\begin{df}
Suppose $\pi$ is a non-degenerate $*$-representation of the Sahl\-mann algebra $\scripta$ on a Hilbert space $\h$. We will say that $\pi$ is an automorphism covariant representation if and only if there exists a family $\{v_\nu\}$ ($\nu\in{\cal N}$) of vectors in $\h$ such that $(i)$ $\h$ admits the orthogonal decomposition $\h=\bigoplus_{\nu\in{\cal N}}\h_\nu$, where each $\h_\nu$ is defined by $v_\nu$ according to (\ref{h-nu}) and $(ii)$ every finite linear combination $v$ of $v_\nu$'s defines an automorphism invariant state on $\scripta$:
\begin{equation}
\scal{v}{\pi(\ttheta\hat{a}\ttheta^{-1})v}\ =\ \scal{v}{\pi(\hat{a})v}
\label{df-inv-eq}
\end{equation}
for every  $a\in \scripta$, and every automorphism $\theta\in\Theta^\omega$.
\label{df-inv}
\end{df}

The consequences of the above definition are the following. Vectors $\{v_\nu\}$ define an isomorphism $\varphi$, which identifies spaces $\bigoplus_{\nu\in{\cal N}}L^2(\Abar,\mu_\nu)$ and $\h$. One can easily check that in terms of this identification the linear subspace ${\cal D}:={\cal C}^\infty$ of $\h$ satisfies the requirements of the Definition \ref{repr-df}. This means that Characterization \ref{sahl-th} is applicable to every automorphism covariant $*$-representation of the Sahlmann algebra.

The requirement \eqref{df-inv-eq} implies that the map
\[
\pi(\hat{\Phi})v_\nu\mapsto \pi(\ttheta\hat{\Phi}\ttheta^{-1})v_\nu,
\]
(where $\Phi\in\cyl^\infty$) defined on $\h_\nu$ is closable and its closure $u_\theta$ is a unitary map on $\h_\nu$. This allow us to conclude that the measure $\mu_\nu$  on $\Abar$ defined by (\ref{measure-df}) is {\em automorphism invariant}, i.e.:
\begin{equation}
\int_{\Abar}\Psi\,d\mu_\nu=\int_{\Abar}(u_\theta\Psi)\,d\mu_\nu.
\label{inv-measure}
\end{equation}
Because of the unitarity of $u_\theta$ the map
\[
\h\ni\Psi\mapsto U_\theta \Psi:=(u_\theta\Psi_\nu)\in\h
\]
defines  a unitary representation of $\Theta^\omega$ on $\h$:
\[
\theta\mapsto U_{\theta}.
\]

Finally, for every $\hat{a}\in\scripta$ the equation
\[
\pi(\ttheta\hat{a}\ttheta^{-1})=U_\theta\pi(\hat{a})U^{-1}_\theta
\]
 is satisfied on ${\cal D}$. Indeed, given $\theta\in\Theta^\omega$, every element $v$ of ${\cal D}$ can be written as 
\[
v=\sum_\nu \pi(\ttheta\hat{\Psi}_\nu\ttheta^{-1})v_\nu.
\]
Hence for any $v,v'\in{\cal D}$ 
\begin{multline*}
\scal{v}{\pi(\ttheta\hat{a}\ttheta^{-1})v'}=\sum_{\nu\nu'}\scal{\pi(\ttheta\hat{\Psi}_{\nu}\ttheta^{-1})v_{\nu}}{\pi(\ttheta\hat{a}\ttheta^{-1})\pi(\ttheta\hat{\Psi}'_{\nu'}\ttheta^{-1})v_{\nu'}}=\\=\sum_{\nu\nu'}\scal{v_{\nu}}{\pi(\ttheta\hat{\Psi}^{*}_{\nu}\hat{a}\hat{\Psi}'_{\nu'}\ttheta^{-1})v_{\nu'}}=\sum_{\nu\nu'}\scal{\pi(\hat{\Psi}_{\nu})v_{\nu}}{\pi(\hat{a})\pi(\hat{\Psi}'_{\nu'})v_{\nu'}}=\\=\scal{v}{U_\theta\pi(\hat{a})U^{-1}_\theta v'}
\end{multline*}
On ${\cal D}$ we also have (see Lemma \ref{lm-transf}):
\begin{equation}
U_\theta\hat{\mathbf{X}}_{S,\lambda}U^{-1}_\theta=\hat{\mathbf{X}}_{\check{\theta}(S),{\rm ad}_\theta\lambda},\;\;\; U_\theta\hat{F}_{S,\lambda}U^{-1}_\theta=\hat{F}_{\check{\theta}(S),{\rm ad}_\theta\lambda}.
\label{diff-hF}
\end{equation}

\section{Main theorem}

Given a $*$-representation $\pi$ of the Sahlmann algebra $\scripta$ on a Hilbert space $\h$, every decomposition (\ref{A}) of $\h$ is determined by a choice of the vectors $\{v_\nu\}$ ($\nu\in{\cal N}$) defining spaces $\h_\nu$ according to (\ref{h-nu}). On the other hand, the representation $\pi$ and the vectors $\{v_\nu\}$ define the family $\{\mu_\nu\}$ of measures  on the space of generalized connection $\Abar$ by means of the formula (\ref{measure-df}).

\begin{thr}
Suppose $\pi$ is a non-degenerate, automorphism covariant $*$-representation of the  Sahl\-mann  algebra
$\scripta$. Then for every member of the family $\{v_\nu\}$ ($\nu\in {\cal N}$) of vectors  satisfying the requirements of Definition \ref{df-inv} the corresponding measure
\[
\mu_\nu\ =\ \mu_{\rm AL},
\]
 where $\mu_{\rm AL}$ is a (defined below) natural measure on $\Abar$. In the consequence, all the $L^2(\Abar,\mu_{\rm AL})$ functions ${{F_{S,\lambda}}^\nu}_\nu$ (no summation) used in Characterization
\ref{sahl-th} are real valued.
\label{main}
\end{thr}

\noindent{\bf Remarks}
\begin{enumerate}
\item The theorem is a generalization of the theorem 5.1 in \cite{ol}. The difference between the two theorems is that in \cite{ol} we assumed $\Sigma=\R^d$ and $P=\Sigma\times G$, while the present theorem is valid for a Sahlmann algebra built on an arbitrary principal bundle $P(\Sigma,G)$ of a compact connected structure group $G$. Moreover, in \cite{ol} we used the notion of a {\em diffeomorphism} covariant representation of the Sahlmann algebra, which is not natural while working with an arbitrary (non-trivial) bundle $P(\Sigma,G)$.
\item Upon the requirement of non-degeneracy and the  definition of the automorphism covariance, every $*$-representation $\pi$ of the Sahlmann $*$-algebra satisfies the conclusions of the theorem.
\end{enumerate}

To define $\mu_{\rm AL}$ let us recall that any trivialization of a bundle $P_{V(\g)}$ defines a bijection between $\A_\g$ and $G^N$, where $N$ is the number of the edges of the graph $\g$. The bijection allows us to push forward the (normalized) Haar measure from $G^N$ onto $\A_\g$. Denote the resulting measure by $\mu_{\g}$. It is easy to show that $\mu_{\g}$ does not depend on the choice of the trivialization of the bundle $P_{V(\g)}$. Let $\psi:\A_\g\rightarrow \C$ be a continuous function. The measure $\mu_{\rm AL}$ is the only measure on $\Abar$ satisfying the requirement \cite{al-meas}
\[
\int_{\Abar} (\bar{p}^*_\g\psi)\, d\mu_{\rm AL}\ = \ \int_{\A_\g} \psi\, d\mu_{\g},
\]
where $\bar{p}_\g:\Abar\rightarrow\A_\g$ is a projection defined analogously\footnote{The projection $\bar{p}_\g$ is defined by an equivalence relation on $\Abar$ analogous to the relation (\ref{rel}) on $\A$. Denote by $\bar{A}_\g$ the equivalence class of a generalized connection $\bar{A}$ i.e., $\bar{A}_\g=:\bar{p}_\g(\bar{A})$. Consequently we denote $\Abar_\g:=\bar{p}_\g(\Abar)$. In fact, every equivalence class $\bar{A}_\g$ contains a smooth connection $A\in\A$ \cite{al-meas}, hence $\Abar_\g$ is naturally isomorphic to $\A_\g$. This justifies the expression $\bar{p}_\g:\Abar\rightarrow\A_\g$ used above.} to the projection $p_\g$ (see Subsection \ref{cyl-funs}).

\section{Proof of the main theorem}

We  have already concluded that  Characterizations \ref{dec} and \ref{sahl-th} are applicable to every $*$-representation $\pi$ satisfying the assumptions of the main theorem. According to the latter characterization the operator $\pi(\hat{X}_{S,\lambda})$ has the following form on ${\cal C}^\infty$:
\begin{equation}
\pi(\hat{X}_{S,\lambda})=\hat{\mathbf{X}}_{S,\lambda}+\hat{F}_{S,\lambda}.
\label{sahl-res}
\end{equation}
It turns out that the automorphism covariance of the representation imposes a restriction on the operator $\hat{F}_{S,\lambda}$ in (\ref{sahl-res})---we will show that there exists a submanifold $C$ and a one-parameter group $\Lambda_\tau$ of vertical automorphisms  of $P_C$ such that for every $\nu\in{\cal N}$ the imaginary part of ${F_{C,\Lambda}}^\nu\!_\nu$ is equal to zero. This fact and Equation (\ref{div-X}) imply that the operator $\hat{X}_{C,\Lambda}$ is {\em symmetric} on $\cyl^\infty\subset L^2(\Abar,\mu_\nu)$. Then we will show that the set of symmetric flux operators is rich enough to use Lemma 6.3 of \cite{ol} to conclude that $\mu_\nu=\mu_{\rm AL}$ for every $\nu\in{\cal N}$.

\subsection{The imaginary part of ${F_{C,\Lambda}}^\nu\!_\nu$}

\subsubsection{The functions ${F_{S,\lambda}}^\iota\!_\nu$}

Consider the functions  ${F_{S,\lambda}}^\iota\!_\nu$ and their imaginary parts:
\[
{I_{S,\lambda}}^\iota\!_\nu:=\frac{1}{2}({F_{S,\lambda}}^\iota\!_\nu-{\overline{F}_{S,\lambda}}^\iota\!_\nu)
\in L^2(\Abar,\mu_\nu).
\]

The assumed automorphism covariance of the representation $\pi$ allows us
to make use of the results derived in Subsection \ref{work}. Equation (\ref{diff-hF}) implies
\[
U_\theta\hat{F}_{S,\lambda}\Psi=\hat{F}_{\check{\theta}(S),{\rm ad}_\theta\lambda}U_\theta\Psi,
\]
for every automorphism $\theta\in\Theta^\omega$ and for every $\Psi\in{\cal C}^\infty$. This means that the
elements ${F_{S,\lambda}}^\iota$ of  $\h$ are assigned
to the submanifolds $S$ and to the family $\lambda_\tau$ in a covariant way,
\begin{equation}
u_\theta{F_{S,\lambda}}^\iota\!_\nu={F_{\check{\theta}(S),{\rm ad}_\theta\lambda}}^\iota\!_\nu, \ \
{\rm hence}\ \
u_\theta{I_{S,\lambda}}^\iota\!_\nu={I_{\check{\theta}(S),{\rm ad}_\theta\lambda}}^\iota\!_\nu.
\label{I-cov}
\end{equation}

The automorphism covariance of the representation $\pi$ also implies
that, for every $\nu$
the scalar product $\scal{\cdot}{\cdot}_\nu$ on $L^2(\Abar,\mu_\nu)$ is
automorphism
invariant, thus in particular
\begin{equation}
||{I_{S,\lambda}}^\iota\!_\nu||_\nu=||u_\theta{I_{S,\lambda}}^\iota\!_\nu||_\nu=||{I_{\check{\theta}(S),{\rm ad}_\theta\lambda}}^\iota\!_\nu||_\nu.
\label{square}
\end{equation}

Equation (\ref{F-sum}) allows us to conclude that for $S=S_1\cup S_2$, where $S_1$
and $S_2$ are  disjoint
\begin{equation}
{I_{S,\lambda}}^\iota\!_\nu={I_{S_1,\lambda_1}}^\iota\!_\nu + {I_{S_2,\lambda_2}}^\iota\!_\nu
\label{sum-I}
\end{equation}
(here $(\lambda_i)_\tau:=\lambda_\tau|_{P_{S_i}}$ and $i=1,2$).


\subsubsection{The functions ${I_{C,\Lambda}}^\nu\!_\nu$ \label{cubes}}

Let us fix the index $\nu$ and denote $I_{S,\lambda}={I_{S,\lambda}}^\nu\!_\nu$ in order to simplify
notation. Recall that $I_{S,\lambda}\in L^2(\Abar,\mu_\nu)$.

Consider a quintuplet
\[
(C,c,\chi,\omega_s,\lambda_\tau)
\]
where $C,c$ are $(d-1)$-dimensional submanifolds\footnote{Precisely speaking we require $C$ and $c$ to be $(d-1)$-dimensional {\em manifolds with boundary} embedded in $\Sigma$. In the sequel of the proof we will associate with $C$ and $c$ some flux operators. Note that although we have defined flux operators only for submanifolds the definition can be extended in a natural way to involve more general objects---consider submanifolds  $S'\subset S$ and $\lambda'_\tau=\lambda_\tau|_{P_{S'}}$. Then $\hat{X}_{S,\lambda}-\hat{X}_{S',\lambda'}$ is a flux operator associated with $S\setminus S'$ being an embedded manifold with boundary. This justifies the existence of the flux operators associated with $C$ and $c$.}  of $\Sigma$, $\chi$ is an element of $\Theta^\omega$,  $\omega_s$ ($s\in\R$) is a family of automorphisms belonging to $\Theta^\omega$ and $\lambda_\tau$ is a one-parameter group of vertical automorphism of $P_c$, (where $P_c$ is the restriction of $P$ to $c$). Assume that the members of the quintuplet satisfy the following conditions:
\begin{enumerate}
\item $C\cup c=\check{\chi}(c)$ and $C\cap c=\varnothing$;
\item every $\check{\omega}_s$ preserves $c$, i.e. for every $s\in \R$
\[
\check{\omega}_s(c)=c;
\]
\item for every point $y\in C$ there exists $s_y\in\R$ such that $s>s_y$ implies $\check{\omega}_s(y)\not\in C$.
\item automorphisms $\chi$ and $\omega_s$ preserve the family $\lambda_\tau$ i.e. for every $\tau,s\in\R$ the following vertical automorphisms on $P_c$ are equal to each other:
\begin{equation}
(\ {\rm ad}_\chi\lambda_\tau\ )|_{P_c}=\lambda_\tau={\rm ad}_{\omega_s}\lambda_\tau
\label{lamb-trans}
\end{equation}
\end{enumerate}
Define now the following family of vertical automorphisms of $P_C$:
\begin{equation}
\Lambda_\tau:=(\ {\rm ad}_\chi\lambda_\tau\ )|_{P_C}.
\label{lambdas}
\end{equation}
We are going  to find a relation between  $I_{c,\lambda}$ and $I_{C,\Lambda}$. Equation (\ref{square}) gives:
\[
||I_{\check{\chi}(c),{\rm ad}_\chi\lambda}||^2_\nu=||I_{c,\lambda}||^2_\nu.
\]
$\check{\chi}(c)=C\cup c$ and $c\cap C=\varnothing$, thus by virtue of Equation (\ref{sum-I}):
\[
I_{\check{\chi}(c),{\rm ad}_\chi\lambda}=I_{c,\lambda}+I_{C,\Lambda}
\]
(note that in the above equation we used the first equality of Equation (\ref{lamb-trans})). Combining the two latter equations we obtain:
\begin{equation}
||I_{C,\Lambda}||^2_\nu=-2\scal{I_{c,\lambda}}{I_{C,\Lambda}}_\nu.
\label{square-I}
\end{equation}

Our goal now is to show that the scalar product $\scal{I_{c,\lambda}}{I_{C,\Lambda}}_\nu$ is equal to 0. We will do it in two steps: first, we will express the function $I_{c,\lambda}$ as a limit of a suitably chosen sequence of cylindrical functions belonging to
$\Cyl^\infty\subset L^2(\Abar,\mu_\nu)$. Then, Characterization
\ref{sahl-th} will allow us to conclude that $\scal{I_{c,\lambda}}{I_{C,\Lambda}}_\nu=0$.

\subsubsection{The function $I_{c,\lambda}$ as a limit of a sequence of cylindrical functions}

The function $I_{c,\lambda}$ can be expressed as a limit,
\[
I_{c,\lambda}=\lim_{n\rightarrow\infty}\Phi_n;\;\;\;\Phi_n\in\cyl^{\infty}.
\]
Note that, in fact, we have quite a large freedom in the choice of the
sequence converging to $I_{c,\lambda}$---to see this, consider a
sequence of automorphisms
$(\theta_n)$ such that $(i)$ every  $\check{\theta}_n$ is analytic and preserves the submanifold $c$ and $(ii)$ every $\theta_n$
preserves the family $\lambda_\tau$ (i.e. ${\rm ad}_{\theta_n}\lambda_\tau=\lambda_\tau$). Then, applying  (\ref{I-cov}) and
the unitarity of $u_\theta$ we obtain:
\[
||I_{c,\lambda}-\Phi_n||_\nu=||u_{\theta_n}(I_{c,\lambda}-\Phi_n)||_\nu=||I_{c,\lambda}-u_{\theta_n}(\Phi_n)||_\nu,
\]
which means that $\lim_{n\rightarrow\infty}u_{\theta_n}(\Phi_n)=I_{c,\lambda}$
as well. We will use this freedom to construct some special
sequence, which converges to $I_{c,\lambda}$.

Let us fix $n$ and consider a graph $\gamma_n$ compatible with
the cylindrical function $\Phi_n$. In general, some edges of the graph can be
transversal\footnote{An edge $e$ is {\em transversal} to a submanifold $S$ iff $S\cap e$ is one-element set. Without  loss of generality $C$ can be expressed as $S\setminus S'$ for some submanifolds $S$ and $S'$. Now, an edge $e$ is by definition {\em transversal} to $C$ if $e$ is transversal to $S$ and $S\cap e=C\cap e$. An appropriate subdivision of each edge transversal to $C$ gives edges of the class $(iii)$ (Subsection \ref{flux-sec}) with respect to submanifold $S$. Thus only transversal edges can give non-zero terms contributing to the action of $\hat{X}_{C,\Lambda}$.} to the submanifold $C$. Then the action of $\hat{X}_{C,\Lambda}$ on $\Phi_n$ is (in general)
nontrivial,
\[
\hat{X}_{C,\Lambda}\Phi_n\neq 0.
\]

\begin{lm}
For each of the graphs $\gamma_n$, $n=1,2,\ldots$ defined above,
there exists an automorphism $\theta_n\in\Theta^\omega$,
such that:
\begin{enumerate}
\item $\check{\theta}_n$ preserves the submanifold $c$ and the action ${\rm ad}_{\theta_n}$ preserves the family $\lambda_\tau$;
\item the graph  $\check{\theta}_n(\gamma_n)$ has no edges transversal to the submanifold $C$.
\end{enumerate}
\label{diff-lm}
\end{lm}
\noindent {\bf Proof.} Fix $n$ and denote by $\{y_1,\ldots,y_M\}$ the set of the intersection points between $C$ and edges of $\gamma_n$ transversal to $C$. The properties of the quintuplet $(C,c,\chi,\omega_s,\lambda_\tau)$ imply the existence of a set $\{s_1,\ldots,s_M\}$ of real numbers such that:
\[
\check{\omega}_{s_J}(y_J)\not\in C.
\]
Let $s(n):={\rm max}\{s_1,\ldots,s_M\}$. Then
\[
\theta_n:= \omega_{s(n)}
\]
preserves $c$ and the family $\lambda_\tau$ and thus satisfies the requirements of the lemma. $\blacksquare$

Use now the automorphisms $(\theta_n)$ given by Lemma \ref{diff-lm}
to construct the following  sequence convergent to $I_{c,\lambda}$,
\[
\tilde{\Phi}_n:=u_{\theta_n}(\Phi_n).
\]
Now, each function in the sequence $(\tilde{\Phi}_{n})$ is compatible with
a graph having no edge transversal to the submanifold $C$, hence
\begin{equation}
\hat{X}_{C,\Lambda}\tilde{\Phi}_n=0.
\label{Xphi}
\end{equation}

\subsubsection{The vanishing of $\scal{I_{c,\lambda}}{I_{C,\Lambda}}_\nu$\label{scal-sec}}

Equation (\ref{div-X}) in the case  $\Phi'=1$ gives
\[
\scal{\hat{X}_{C,\Lambda}\Phi}{1}_\nu=2\scal{\Phi}{I_{C,\Lambda}}_\nu,
\]
for every smooth cylindrical function $\Phi\in L^2(\Abar,\mu_\nu)$. Owing to
Equation (\ref{Xphi})
\[
0=\lim_{n\rightarrow\infty}\scal{\hat{X}_{C,\Lambda}
\tilde{\Phi}_n}{1}_\nu=\lim_{n\rightarrow\infty}2\scal{\tilde{\Phi}_n}
{I_{C,\Lambda}}_\nu=2\scal{I_{c,\lambda}}{I_{C,\Lambda}}_\nu.
\]
This result and Equation (\ref{square-I}) imply
\begin{equation}
||I_{C,\Lambda}||^2_\nu=0, \ \ {\rm hence}\ \   I_{C,\Lambda}={I_{C,\Lambda}}^\nu\!_\nu=0,
\label{I-zero}
\end{equation}
where the last equality refers to elements of $L^2(\Abar,\mu_\nu)$
while the measure $\mu_\nu$ is not assumed to be faithful.

\subsection{Construction of the quintuplet \label{constr}}

To justify the result (\ref{I-zero}) we have to show that there exists a quintuplet $(C,c,\chi,\omega_s,\lambda_\tau)$ satisfying all the requirements described in Subsection \ref{cubes}. This will be done by an explicit construction.

To construct the quintuplet we have to take into account that the automorphisms $\chi$ and $\omega_s$ have to be  elements of $\Theta^\omega$ i.e., that they have to define {\em globally} analytic diffeomorphisms $\check{\chi}$ and $\check{\omega}_s$ on $\Sigma$. On the other hand, the diffeomorphisms have to transform (and, in particular, preserve) submanifolds $C$ and $c$ in the required way. Because it is rather difficult to construct an analytic diffeomorphism which e.g. preserves a given submanifold we will first construct the diffeomorphisms and only then  we will find appropriate submanifolds $C$ and $c$.

The idea of the construction of the quintuplet is the following: $(i)$ first we will find an analytic vector field $Y$ on $\Sigma$ such that it generates one-parameter group of analytic diffeomorphisms on $\Sigma$; $(ii)$ the diffeomorphisms  $\check{\chi}$ and $\check{\omega}_s$ will be defined as maps which shift points along the integral curves of $Y$, i.e. preserve the curves; $(iii)$ the submanifolds $C$ and $c$ will be constructed  by the shift of a $(d-2)$-dimensional submanifold along the integral curves of $Y$; $(iv)$ we will obtain automorphisms $\chi$ and $\omega_s$ by a horizontal lift (with respect to a connection on $P$) of $\check{\chi}$ and $\check{\omega}_s$; $(v)$ finally we will use a section of the bundle $P_c$ to define $\lambda_\tau$.


\subsubsection{A construction of analytic diffeomorphisms}

It is well known that every (sufficiently regular) vector field on a manifold generates a {\em local} one-parameter group of {\em local} diffeomorphisms on this manifold (see e.g. \cite{kobayashi}), which in the case of non-compact manifolds is not always extendable to a group of diffeomorphisms. Therefore to obtain (a one-parameter group of) diffeomorphisms on $\Sigma$ we have to choose the generating vector field in a suitable way.  To choose it we will apply a theorem by Grauert \cite{gra}, which states that every analytic manifold can be embedded into $\R^k$ of sufficiently high dimension by means of an {\em analytic proper} map. Because the properness of the map will be crucial for the below considerations let us recall the definition of the notion.

Let $T_1,T_2$ be topological spaces. A continuous map $\xi: T_1\rightarrow T_2$ is proper  if and only if  for every compact $U\subset T_2$ the inverse image $\xi^{-1}(U)$ is compact in $T_1$. One can show in particular that $\xi(T_1)$ is closed in $T_2$  and if $\xi$ is injective, then $\xi$ is homeomorphism between $T_1$ and $\xi(T_1)$ with topology induced by the topology of $T_2$ (see e.g. \cite{bourbaki}). 

Let us  assume then that $\Sigma$ is embedded into some $\R^k$ by an analytic proper map. Denote by $q$ the Riemannian metric on $\Sigma$ induced by the Euclidean metric on $\R^k$. Then the following lemma holds:

\begin{lm}
Suppose $Y$ is an analytic vector field on $\Sigma$ such that the function $q(Y,Y)$ is bounded on $\Sigma$. Then $Y$ generates a one-parameter group $\check{\alpha}_t$ of  analytic diffeomorphisms on $\Sigma$.
\label{complet}
\end{lm}
Before we prove the lemma let us note that the lemma is not true for an arbitrary Riemannian metric on $\Sigma$, as it is shown by the following example:

\begin{eg}
\rm Let $\Sigma=\R$ and $g(\partial_x,\partial_x)=\exp(-x^2)$, then for $Y=x^2\partial_x$ function $g(Y,Y)$ is bounded on $\R$, but $Y$ does not generate a group of diffeomorphisms on $\R$. $\blacksquare$
\label{Rq}
\end{eg}

\noindent{\bf Proof of Lemma \ref{complet}.} Every analytic vector field generates  a local one-parameter group of local {\em analytic} diffeomorphisms (see e.g. \cite{maurin}). Thus to prove the lemma it is enough to show that every {\em non-extendable} integral curve of $Y$ is a map from the {\em whole} $\R$ into $\Sigma$. Let us consider then a non-extendable integral curve:
\begin{equation}
\R\supset ]a,b[ \ni t\mapsto y(t)\in \Sigma.
\label{t-curve}
\end{equation}
where $a,b$ can be equal to $\pm\infty$ respectively.

The theory of ODE's (see e.g. \cite{maurin}) guarantees that if $Y(y(t_0))=0$ for some $t_0\in]a,b[$, then $y(t)\equiv y(t_0)$ and consequently $]a,b[=\R$. If a given integrable curve forms a loop, then obviously: $]a,b[=\R$.

Now let us suppose that $Y(y(t))\neq 0$ on $]a,b[$ and the integral curve does not form a loop. Parametrize the path $y(]a,b[)$ by its length $\tau$ defined by the metric $q$. Thus after reparametrization of the curve $y(t)$ we get a curve:
\begin{equation}
\R\supset ]a',b'[ \ni \tau\mapsto y'(\tau)\in \Sigma.
\label{tau-curve}
\end{equation}
We have:
\[
Y(y'(\tau))=\frac{d\tau}{dt}\partial_\tau, \ \ \text{hence} \ \ q(Y,\partial_\tau)=\frac{d\tau}{dt}> 0
\]
(here $\partial_\tau$ is a unital vector field on the submanifold $y(]a,b[)$ generated by the curve $y'(\tau)$.) Integrating the latter equation we obtain:
\begin{equation}
t(\tau)=\int_{\tau_0}^{\tau}\frac{d\tau'}{q(Y,\partial_\tau)},
\label{int}
\end{equation}
where $\tau_0$ is chosen in a way satisfying $t(a')=a$ and $t(b')=b$.

Assume that $b'=\infty$ ($a'=-\infty$). Because $q(Y,\partial_\tau)>0$ and is bounded (by virtue of the Schwarz inequality and the boundedness of $q(Y,Y)$) the l.h.s. of Equation (\ref{int}) is not bounded from above (below), hence $b=\infty$ ($a=-\infty$).

Suppose now that $b'<\infty$ and $b=t(b')<\infty$. Then:
\begin{equation}
\lim_{t\rightarrow b}y(t)=\lim_{\tau\rightarrow b'}y'(\tau)=\mathbf{y}\in\Sigma.
\label{limit}
\end{equation}
Actually, the existence of the limit (\ref{limit}) is a consequence of the fact that the metric $q$ is induced by the Euclidean metric\footnote{If $q$ was an arbitrary metric on $\Sigma$, then $\lim_{\tau\rightarrow b'}y'(\tau)$ could not exist --- consider $\Sigma$ and $g$ as in Example \ref{Rq}. Then $\Sigma=\R$ can be parametrized by length parameter $]a',b'[\ \ni\tau\mapsto y'(\tau)\in \Sigma$, where $a',b'$ are finite. Clearly $\lim_{\tau\rightarrow b'}y'(\tau)=\infty$.} on $\R^k$, hence we can consider the curve $y'(\tau)$ as a curve in $\R^k$ and find the limit in this space. Because $\Sigma$ is a closed subset of $\R^k$ (which follows from the properness of the map embedding $\Sigma$ in the $\R^k$) the limit belongs to $\Sigma$.

The theory of ODE's applied to the {\em analytic} vector field $Y$ ensures that there exists a neighbourhood $U\subset\Sigma$ of $\mathbf{y}$ and $\delta>0$ such that for every $(t_0,y_0)\in\ ]-\delta,\delta[\ \times U$ there exists {\em exactly one} curve $\tilde{y}:\ ]-\delta,\delta[\ \rightarrow\Sigma$ being an integral curve of $Y$ and satisfying the initial condition $\tilde{y}(t_0)=y_0$. The assumption $b<\infty$ implies that 
\begin{enumerate}
\item if $Y(\mathbf{y})\neq 0$ then the curve $y(t)$ under consideration is extendable, which contradicts our assumption about the curve,
\item If $Y(\mathbf{y})=0$ then there exist two {\em distinct} solutions of ODE $dy/ds=Y$ with the initial value $y(0)=\mathbf{y}$: one of them is $y(s)\equiv \mathbf{y}$ and the second one is:
\[
y(s)=
\begin{cases}
y(s)=y'(\tau(s))&\text{for $s<0$}\\
y(s)=\mathbf{y}&\text{for $s\geq0$}
\end{cases},
\]
where $\tau(s)$ is given by $\int^{b'}_{\tau(s)}d\tau/q(Y,\partial_\tau)=-s$. Clearly, this is forbidden by the theory of ODE's as written above.
\end{enumerate}
In this way we conclude that $b=\infty$. Similarly we have $a=-\infty$ even if $a'> -\infty$. Thus $]a,b[=\R$. $\blacksquare$

\begin{eg}
\rm Let us fix a point $y_0\in\Sigma$ and a nonzero vector $Y_{y_0}\in T_{y_0}\Sigma$ and suppose for simplicity that $\Sigma$ is embedded into some $\R^k$ in such a way that $y_0$ is placed at the origin of the Cartesian coordinate frame of $\R^k$. Clearly, $Y_{y_0}$ defines a constant vector field on $\R^k$. Let $\tilde{Y}$ be a vector field on $\R^k$ given by the following formula:
\[
\tilde{Y}(x):=\exp(-\|x\|^2_\R)Y_{y_0},
\]
where $x\in\R^k$ and $\|\cdot\|_\R$ is the standard norm on $\R^k$. Let $Y$ be a vector field on $\Sigma$ defined as the orthogonal (with respect to the standard metric of $\R^k$) projection of $\tilde{Y}$ onto the tangent bundle $T\Sigma$. Then $Y$ is an analytic vector field on $\Sigma$ satisfying the assumption of Lemma \ref{complet} (and condition $Y(y_0)=Y_{y_0}$). Hence $Y$ generates a one-parameter group $\check{\alpha}_t$ of analytic diffeomorphisms on $\Sigma$. $\blacksquare$ \label{vector-Y}
\end{eg}

In this way we finished the step $(i)$ of the construction.

\subsubsection{Construction of the quadruplet $(C,c,\check{\chi},\check{\omega}_s)$}

Now let us construct the quadruplet $(C,c,\check{\chi},\check{\omega}_s)$ satisfying the first three conditions described in Subsection \ref{cubes} (steps $(ii)$ and $(iii)$ of the construction).

Fix an arbitrary point $y_0\in\Sigma$ and consider the following ingredients, from which the quadruplet $(C,c,\check{\chi},\check{\omega}_s)$ will be built:
\begin{enumerate}
\item Let $Y$ be an analytic  vector field on $\Sigma$ satisfying assumption of Lemma \ref{complet} such that $Y(y_0)\neq 0$. Denote by $\check{\alpha}_t$ the one-parameter group of diffeomorphisms generated by $Y$.
\item Let $\sigma$ be a $(d-1)$-dimensional analytic (possibly non-closed) submanifold of $\Sigma$ such that $y_0\in\sigma$ and $Y(y_0)$ is {\em not} tangent to $\sigma$.
\item Let $\xi:\Sigma\rightarrow \R$ be a bounded analytic function such that for some neighbourhood $\tilde{U}$ of $y_0$ $(i)$ $\xi=0$ on $\tilde{U}\cap\sigma$ and $(ii)$ $\xi > 0$ on $\tilde{U}\setminus\sigma$ (such a function can be easily constructed by means of the Grauert theorem --- see Example \ref{eg-quadr}).
\end{enumerate}

Then there exists a neighbourhood $u\subset\sigma$ of point $y_0$, an (analytic) coordinate frame $(y^1,\ldots,y^{d-1})$ on $u$ and $t_0>0$ such that:
\begin{enumerate}
\item $(t,y^i)$ is an analytic coordinate frame on:
\[
U:=\{\check{\alpha}_t(u)\ | \ t\in\ ]-3t_0,3t_0[\ \}\subset\tilde{U}\subset\Sigma
\]
\item for every $y\in U$:
\begin{equation}
Y(y)\neq 0.
\label{YUnot0}
\end{equation}
\end{enumerate}
Now let us define a function $\xi'$ as the pull-back of $\xi$:
\[
\xi':=\check{\alpha}^*_{t_0}\xi
\]
and the two vector fields satisfying assumption of Lemma \ref{complet} as
\begin{equation}
Z:=\xi'Y\ \ \text{and} \ \ W:=\xi\xi'Y.
\label{ZW}
\end{equation}
Denote by $\check{\beta}_s$ and $\check{\omega}_s$ the one-parameter group of diffeomorphisms generated by the vector fields $Z$ and $W$ respectively.

\begin{figure}
\psfrag{a+u}{$\check{\alpha}_{t_0}(u)$}
\psfrag{a+}{$\check{\alpha}^+(u)$}
\psfrag{s}{$\sigma$}
\psfrag{u}{$u$}
\psfrag{Y}{$Y(y_0)$}
\psfrag{a-}{$\check{\alpha}^-(u)$}
\psfrag{a-u}{$\check{\alpha}_{-t_0}(u)$}
\psfrag{bsu}{$\check{\beta}_{s_0}(u)$}
\psfrag{C}{$C$}
\psfrag{cu}{$c_u$}
\psfrag{c}{$c$}
\psfrag{atu}{$\check{\alpha}_{-t_0}(c_u)$}
\centerline{\includegraphics{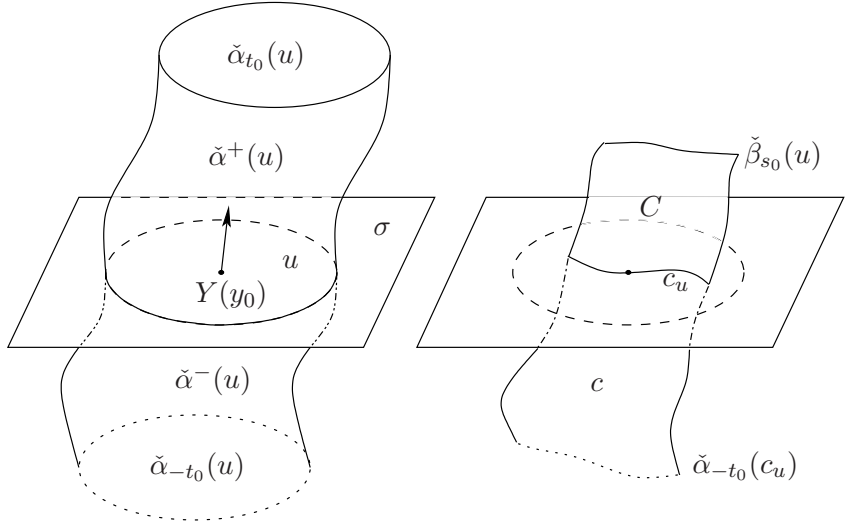}}
\caption{Construction of submanifolds $C$ and $c$ \label{picture}}
\end{figure}

Consider now sets (see Figure \ref{picture}):
\begin{align*}
\check{\alpha}(u):=&\{ \check{\alpha}_t(u)\ | \ t\in\ ]-t_0,t_0]\ \}\\
\check{\alpha}^+(u):=&\{ \check{\alpha}_t(u)\ | \ t\in \ ]0,t_0]\ \}\\
\check{\alpha}^-(u):=&\{ \check{\alpha}_t(u)\ | \ t\in\ ]-t_0,0]\ \}
\end{align*}
(we emphasize that $\check{\alpha}^-(u)$ contains $u$; clearly $\check{\alpha}(u)=\check{\alpha}^+(u)\cup\check{\alpha}^-(u)$). The properties of the function $\xi$ guarantee that $\xi'$ is {\em non-zero} on $\check{\alpha}(u)$ and that it is equal to zero on $\check{\alpha}_{-t_0}(u)$. Taking into account Equation (\ref{YUnot0}), one can easily conclude that if $s\neq 0$, the diffeomorphism $\check{\beta}_s$ is not the identity on the whole $\check{\alpha}(u)$ and it is the identity on $\check{\alpha}_{-t_0}(u)$ for every $s\in\R$. Similarly, for $s\neq0$ diffeomorphism $\check{\omega}_s$ is not the identity on $\check{\alpha}(u)\setminus u$ and is the identity on $\check{\alpha}_{-t_0}(u)$ and $u$.

Except for the points where vector fields $Z$ and $W$ vanish, the integral curves of $Y,Z$ and $W$ define the same paths. Thus every $\check{\omega}_s$ preserves $\check{\alpha}^-(u)$. Moreover, for every point $y\in\check{\alpha}^+(u)$ there exists $s_y\in\R$ such that $s>s_y$ implies\footnote{Note that $\check{\alpha}^+(u)$ does {\em not} contain any points, at which $\check{\omega}_{s\neq0}$ is the identity. Moreover, $u$ is the only subset of the boundary of $\check{\alpha}^+(u)$, where $\check{\omega}_s={\rm id}$. Thus if $s_y$ is large enough $\check{\omega}_{s_y}$ shifts a given point $y\in\check{\alpha}^+(u)$ out this set. Note as well that because $\check{\omega}_s$ is the identity on $u$ the integral curves of $W$ passing through $\check{\alpha}^+(u)$ do not form loops---hence $\check{\omega}_s(y)\not\in\check{\alpha}^+(u)$, only if $s>s_y$.} $\check{\omega}_s(y)\not\in\check{\alpha}^+(u)$.

Now we are ready to define the quadruplet $(C,c,\check{\chi},\check{\omega}_s)$. In fact, $\check{\omega}_s$ is already defined as a family of diffeomorphisms generated by $W$. It is possible to choose $(i)$ $(d-2)$-dimensional regular analytic submanifold\footnote{If $\dim\Sigma=2$ take $c_u=y_0$.} $c_u$ contained in $u$ and  containing the point $y_0$ and $(ii)$ a positive number $s_0$ such that the following analytic $(d-1)$-dimensional submanifold
\[
C:=\{\check{\beta}_{s}(c_u)\ | \ s\in \ ]0,s_0]\}
\]
is contained in $\check{\alpha}^+(u)$. Define:
\[
c:=\{\check{\alpha}_t(c_u)\ | \ t\in \ ]-t_0,0]\}
\]
(clearly, $c$ is contained in $\check{\alpha}^-(u)$) and:
\begin{equation}
\check{\chi}:=\check{\beta}_{s_0}.
\label{chchi-df}
\end{equation}
Hence we have $C\cup c=\check\chi(c)$ and $C\cap c=\varnothing$. In this way we completed the construction of  the quadruplet $(C,c,\check{\chi},\check{\omega}_s)$ satisfying the first three requirements described in Subsection \ref{cubes}.

\begin{eg}
\rm (continuation of Example \ref{vector-Y}) To construct a quadruplet $(C,c,\check{\chi},\check{\omega}_s)$ using the vector field $Y$ described in Example \ref{vector-Y} we only need to define the submanifold $\sigma$ and the function $\xi$. Let $(\cdot|\cdot)$ be the standard scalar product on $\R^k$. Define the function
\[
\R^k\ni x\mapsto \tilde{\xi}(x):=1-\exp[-(x|Y(y_0))^2]\in\R.
\]
Clearly $\tilde{\xi}$ is bounded nonnegative function on $\R^k$ and it vanishes on the $(k-1)$-dimensional plane $\mathbb{P}$ orthogonal to the vector $Y(y_0)$. Let $B(r,y_0)$ be a ball in $\R^k$ of center in $y_0$ and radius $r>0$. Thus we can define:
\begin{align*}
\xi:=&\tilde{\xi}|_\Sigma\\
\sigma:=&\mathbb{P}\cap B(r_0,y_0)\cap\Sigma
\end{align*}
for some (small enough) $r_0>0$. $\blacksquare$
\label{eg-quadr}
\end{eg}

\subsubsection{Construction of automorphisms $\chi$, $\omega_s$ and $\lambda_\tau$}

Now, by suitable lifts of diffeomorphisms $\check{\chi}$ and $\check{\omega}_s$ we will get the desired automorphisms $\chi$ and $\omega_s$ (step $(iv)$) and we will then define the family $\lambda_\tau$ of vertical automorphisms on $P_c$ (step $(v)$), which will complet the construction of the quintuplet $(C,c,\chi,\omega_s,\lambda_\tau)$.

Let $T$ denote a neighbourhood of the point $y_0$ (around which submanifolds $C$ and $c$ were just constructed) such that the map
\[
\pr^{-1}(T)\ni p\mapsto\varsigma(p)=(\pr(p),g_p)\in T\times G
\]
 is a local trivialization of the bundle $P$. It is clear that we can construct the manifolds $C,c$ in such a way that
$c\subset T$. Moreover, without loss of generality we can assume that there exist neighbourhoods $T_0,T_1$ of $y_0$ such that $(i)$ $c\subset T_0\subset T_1\subset T$ and $(ii)$ there exists a smooth function $\eta$ on $\Sigma$ equal to $0$ on $T_0$ and $1$ on $\Sigma\setminus T_1$.

Let $A$ be any connection one-form on $P$. Define the connection $A_0$ in the following way:
\[
\begin{cases}
A_0:=A & \text{on $\pr^{-1}(\Sigma\setminus T)$}\\
\varsigma^*_0 A_0:=\eta\ \varsigma^*_0 A & \text{on $T$}
\end{cases}
\]
where $\varsigma_0$ is a local section of $P$ over $T$ defined as $\varsigma(\varsigma_0(y))=(y,\mathbb{I})$ ($y\in T$, $\mathbb{I}$ is the neutral element of $G$). Note that on $T_0$ the pull-back $\varsigma^*_0 A_0=0$, which means that every set $\varsigma^{-1}(y,g={\rm const})$, $y\in T_0$ defines a (local) section of $P$ which is {\em horizontal} with respect to the connection $A_0$.

Let $Z^*$ and $W^*$ be defined by $A_0$ horizontal lifts  of the vector fields $Z$ and $W$ (see Equation (\ref{ZW})), respectively. Denote by $\beta_s$ and $\omega_s$ the one-parameter groups  of diffeomorphisms on $P$ generated by $Z^*$ and $W^*$, respectively. Clearly, the diffeomorphisms $\beta_s$ and $\omega_s$ are automorphisms of $P$ and elements of $\Theta^\omega$. Define now:
\[
\chi:=\beta_{s_0},
\]
where $s_0$ is the same real number as in (\ref{chchi-df}).

Let $f$ be a constant function from $c$ to the Lie algebra $G'$. The desired family $\lambda_\tau$ can be defined using the trivialization $\varsigma$ in the following way:
\begin{equation}
P_c\ni p\mapsto \lambda_\tau(p):=(\pr(p),\exp(\tau f)g_p)\in P_c.
\label{lambda-tau}
\end{equation}
In the trivialization $\varsigma$ the automorphisms $\chi$ and $\omega_s$ act in the following way:
\[
\chi(p)=(\check{\chi}(\pr(p)),g_p),\;\;\;\omega_s(p)=(\check{\omega}_s(\pr(p)),g_p),
\]
provided $p\in\pr^{-1}(T_0)$---this is because, by virtue of the construction, the vector fields $Z^*$ and $W^*$ are tangent to the submanifolds defined by $(\pr(p),$ $ g_p={\rm const})$. Thus it is easy to see that $\lambda_\tau$ satisfies the condition (\ref{lamb-trans}).

This completes the construction of the quintuplet $(C,c,\chi,\omega_s,\lambda_\tau)$.

\subsection{Final conclusion \label{final}}

Lemma 6.2 of \cite{ol} can be reformulated in the following way (see also \cite{oko-phd}):
\begin{lm}
Let $G$ be a compact, connected Lie group and $\mu$ be a probability (regular Borel) measure on $\Abar$. Suppose that for every collection of edges $\{e_1,\ldots,e_N\}$ constituting a graph there exists $N\times\dim G$ operators $X_{S_I,\lambda^i_I}$ $(I=1,\ldots,N$, $i=1,\ldots,\dim G)$ such that $(i)$ the operators $X_{S_I,\lambda^i_I}$ are symmetric on $\cyl^\infty\subset L^2(\Abar,\mu)$ $(ii)$ for every $I$ the intersection $S_I\cap(e_1\cup\ldots\cup e_N)=S_I\cap e_I$ is a one-element set and $e_I$ is transversal to $S_I$ $(iii)$ for every $I$ the families $\{\lambda^i_I\}$ generate a reper field on $P_{S_I\cap e_I}$. Then:
\[
\mu=\mu_{\rm AL}.
\]
\label{lm-al}
\end{lm}

Before we  show that in our case all assumptions of the lemma are satisfied let us make some remarks. The lemma means that for every $\nu\in {\cal N}$ the measure $\mu_\nu=\mu_{\rm AL}$, thus it completes the proof of the first conclusion of the main theorem.
 On the other hand, {\it every} operator $\hat{X}_{S,\lambda}$ is self-adjoint on
 $L^2(\Abar,\mu_{\rm AL})$ \cite{area}. Thus the first conclusion  and Equation (\ref{div-X}) imply the
second conclusion of the theorem i.e. that for every $\nu$ the function ${F_{S,\lambda}}^\nu\!_\nu$ is real valued.

Let us turn back to the assumptions of the lemma. Equation (\ref{div-X}) and (\ref{I-zero}) imply that for $C$ and $\Lambda_\tau$  considered in the last subsection and for every smooth cylindrical functions $\Phi,\Phi'$:
\[
\scal{\hat{X}_{C,\Lambda}\Phi}{\Phi'}_\nu-\scal{\Phi}{\hat{X}_{C,\Lambda}\Phi'}_\nu=0,
\]
which means that $\hat{X}_{C,\Lambda}$ is {\em symmetric} on $\cyl^\infty\subset L^2(\Abar,\mu_\nu)$. Moreover, for every $\theta\in\Theta^\omega$ the operator $\hat{X}_{\check{\theta}(C),{\rm ad}_\theta\Lambda}$ is symmetric on $\cyl^\infty$ as well. Therefore now we only have to show  that there exist sufficiently many operators of the form $\hat{X}_{\check{\theta}(C),{\rm ad}_\theta\Lambda}$, i.e. that the assumptions $(ii)$ and $(iii)$ are satisfied by the operators.

Note that for every edge $e\subset \Sigma$ there exists a submanifold $C$ (being a member of a quintuplet) and an automorphism $\theta\in\Theta^\omega$ such that a submanifold $\check{\theta}(C)$ intersects $e$ at exactly one point not being any vertex of $e$ (this is required in fact by the assumption $(ii)$) and $e$ is transversal to the submanifold. To see this, choose a point $y_0\in {\rm Int}\ e$ and a vector $Y_{y_0}\in T_{y_0}\Sigma$ such that $Y_{y_0}$ is {\em not} tangent to $e$. Applying the construction described in Examples \ref{vector-Y} and \ref{eg-quadr} we obtain a vector field $Y$ with a family of diffeomorphisms $\check{\alpha}_t$ and a quintuplet containing a submanifold $C$---this submanifold can be defined (by an appropriate choice of $(d-2)$-dimensional submanifold $c_u$) in such a way that $\check{\alpha}_{-t_0}(C)$ for some (small enough) $t_0>0$ meets $e$ transversally at exactly one interior point of the edge. Then one can find $\theta\in\Theta^\omega$ such that $\check{\theta}=\check{\alpha}_{-t_0}$. Thus $e$ is transversal to $\check{\theta}(C)$ and the assumption $(ii)$ is satisfied.

Consider now the submanifold $c$ constructed together with the submanifold $C$. It is easy to choose $n=\dim G$ families $\{\lambda^i_\tau\}$ of vertical automorphisms acting on $P_c$ such that they generate a reper field on every fibre of the bundle\footnote{Here the notion of a reper field concerns the fibre thought as a separate manifold.}---indeed, we have constructed the family $\lambda_\tau$ by means of the (constant) function\footnote{$G'$ indicates the Lie algebra of the group $G$.} $f:c\rightarrow G'$ (Equation (\ref{lambda-tau})), therefore $n$ constant functions $\{f^i\}$ define desired families  $\{\lambda^i_\tau\}$ provided the values of these functions form a basis of $G'$. Thus the families
\[
{\rm ad}_\theta\Lambda^i_\tau ={\rm ad}_\theta [(\ {\rm ad}_\chi \lambda^i_\tau \ )|_{P_C}]
\]
generate (in particular) a reper field on the fibre over the intersection point between $\check{\theta}(C)$ and the edge $e$. This means that the assumption $(iii)$ is satisfied.

In this way we have finished the proof of the main theorem.

\section{Summary}

In this paper we considered automorphism covariant
$*$-represen\-ta\-tions of the Sahlmann ho\-lo\-no\-my-flux algebra
$(\scripta,*)$ for a theory of connections on an arbitrary bundle $P(\Sigma,G)$, where $\Sigma$  is an arbitrary real-analytic manifold and $G$ is a compact
connected Lie group. We showed that the carrier space of the $*$-representations used in Sahlmann's Characterization \ref{sahl-th} is
the orthogonal product (Theorem \ref{main}):
\begin{equation}
\h=\bigoplus_\nu L^2(\Abar,\mu_\nu),
\label{sum-hilb}
\end{equation}
where every measure $\mu_\nu$ is the natural measure:
\begin{equation}
\mu_\nu=\mu_{\rm AL}.
\label{nu-al}
\end{equation}
\bigskip

We emphasize that, although the requirement of the automorphism covariance of the representation singles out the measure $\mu_{\rm AL}$, there may exist inequivalent automorphism covariant representations of the Sahlmann algebra on the Hilbert space (\ref{sum-hilb}) which differ from each other by the family of functions ${F_{S,\lambda}}^\iota\!_\nu$ (note that the functions are not fully determined by Theorem \ref{main}). Thus automorphism covariance with respect to the group of automorphisms whose projections are {\em analytic} diffeomorphisms, (i.e. the group $\Theta^\omega$) does not seem to fix a unique automorphism covariant representation.

It is, however, possible to define an action of a larger group of automorphisms on the holonomy-flux $*$-algebra and modify slightly the definition of the automorphism covariance of representations of the algebra. Then all the considerations presented here and Theorem \ref{main} remain valid. We can also go a step further  and prove that the automorphism covariance with respect to the extended group of automorphisms singles out precisely one representation of the Sahlmann algebra  which is the cyclic representation described by Characterization \ref{sahl-th} with all the correction terms equal zero. This result will be presented in \cite{lost}.\bigskip

{\bf Acknowledgements:} Our interest in this subject was stimulated by discussions with Hanno Sahlmann who also explained to us many technical points and discussed with us his unpublished considerations.
Stanis{\l}aw L. Woronowicz encouraged us to reformulate our results in the language of non-trivial principal bundles. We are grateful to Marcin Bobie\'nski for a discussion which helped us to formulate Lemma 6.2. Finally, we thank Abhay Ashtekar and Thomas Thiemann for numerous discussions about the quantum geometry and, respectively, QSD. This work was supported in part by the KBN grants 2 PO3B 068 23 and  2 PO3B 127 24 and by the NSF grant PHY 0090091.

\end{document}